%% file: paperCaenMethodological.tex
\documentclass{article}
\usepackage[inline, draft]{fixme}
\fxsetup{theme=color, mode=multiuser} \FXRegisterAuthor{me}{ame}{\color{\colorComments}Me}
  \usepackage[T1]{fontenc}
  \usepackage[utf8]{inputenc}
  \usepackage{lmodern}
  \DeclareUnicodeCharacter{327}{-}
\usepackage{natbib}
\setcitestyle{authoryear,open={(},close={)}}
\usepackage{rotating}

\usepackage{framed,multirow}
\usepackage{acronym}
\usepackage{booktabs}
\usepackage{graphicx}
\usepackage{subcaption}
\usepackage{placeins}
\usepackage{multicol}
\usepackage{amssymb}
\usepackage{latexsym}

\usepackage{url}
\usepackage{xcolor}
\usepackage{blindtext}
\usepackage[]{algorithm2e}
\usepackage{amsmath}

\usepackage{float}

\usepackage{hyperref}
\definecolor{newcolor}{rgb}{.8,.349,.1}

\usepackage{geometry}

\usepackage{centernot}
\newcommand{\adjacent}[1][.7em]{\mathrel{\rule[.5ex]{#1}{.4pt}}}

\usepackage{authblk}

\usepackage{efbox,graphicx}
\efboxsetup{linecolor=green,linewidth=10pt}

\title{Volumetric parcellation of the cardiac right ventricle for regional geometric and functional assessment.}
\newcommand{\colorComments}{black}
\usepackage{fancyhdr}
\fancypagestyle{plain}{\pagestyle{fancy}}
\begin{document}
\author[1,*]{Bernardino, G}
\author[2,*]{Hodzic, A}
\author[4]{Langet, H}
\author[5]{Legallois, D}
\author[4]{De Craene, M}
\author[1,6]{Gonz\'alez Ballester, MA}
\author[5]{Saloux, E}
\author[6,7,8]{Bijnens, B}

\affil[1]{BCN Medtech, Dept. of Information and Communication Technologies, Universitat Pompeu Fabra, Barcelona, Spain}
\affil[2]{Normandie Univ, UNICAEN, CHU de Caen Normandie, Department of Clinical Physiology, Inserm Comete, GIP Cyceron, Caen, France.}
\affil[4]{Philips Research, Paris, France}
\affil[5]{ Normandie Univ, UNICAEN, CHU de Caen Normandie, Department of Cardiology, EA4650 SEILIRM, Caen, France.}

\affil[5]{ ICREA, Barcelona, Spain}
\affil[6]{ IDIBAPS, Barcelona, Spain}
\affil[7]{ KULeuven, Leuven, Belgium}
\affil[*]{A. Hodzic and G. Bernardino contributed equally to this work}


\newacro{SSA}{Statistical shape analysis}
\newacro{MRI}{ magnetic resonance imaging}
\newacro{SA}{ short axis}

\newacro{GPPA}{ generalized partial procrustes analysis}

\newacro{PDM}{ point distribution model}
\newacro{CV}{ cross validation}

\newacro{DR}{ dimensionality reduction}

\newacro{BMI}{body mass index}
\newacro{BSA}{body surface area}

\newacro{LV}{left ventricle}
\newacro{RV}{right ventricle}
\newacro{ED}{end-diastole}
\newacro{ES}{end-systole}
\newacro{EF}{ejection fraction}
\newacro{HR}{heart rate}
\newacro{RVOT}{right ventricular outflow tract}
\newacro{HCM}{hypertrophic cardiomyopathies}

\maketitle
\begin{abstract}
    \input{abstract.tex}
\end{abstract}
\acresetall
\input{mainTextMEDIAR1.tex}
\bibliography{references}

\end{document}

%% file: abstract.tex
3D echocardiography is an increasingly popular tool for assessing cardiac remodelling in the \ac{RV}. It allows quantification of the cardiac chambers without any geometric assumptions, which is the main weakness of 2D echocardiography. However, regional quantification of geometry and function is limited by the lower spatial and temporal resolution and the scarcity of identifiable anatomical landmarks, especially within the ventricular cavity. We developed a technique for regionally assessing the volume of 3 relevant RV volumetric regions: apical, inlet and outflow.

The proposed parcellation method is based on the geodesic distances to anatomical landmarks that are easily identifiable in the images: the apex and the tricuspid and pulmonary valves, each associated to a region. Based on these distances, we define a partition in the endocardium at \ac{ED}. This partition is then interpolated to the blood cavity using the Laplace equation, which allows to compute regional volumes. For obtaining an \ac{ES} partition, the endocardial partition is transported from \ac{ED} to \ac{ES} using a commercial image-based tracking software, and then the interpolation process is repeated. We assessed the intra- and inter-observer reproducibility using  a 10-subjects dataset containing repeated quantifications of the same images, obtaining intra- and inter- observer errors ($7-12\%$ and $10-23\%$ respectively).

Finally, we propose a novel synthetic mesh generation algorithm that deforms a template mesh imposing a user-defined strain to a template mesh. We used this method to create a new dataset for involving distinct types of remodelling that were used to assess the sensitivity of the parcellation method to identify volume changes affecting different parts. We show that the parcellation method is adequate for capturing local circumferential and global circumferential and longitudinal RV remodeling, which are the most clinically relevant cases.

%% file: mainTextMEDIAR1.tex



\newacro{SSA}{Statistical shape analysis}
\newacro{MRI}{ magnetic resonance imaging}
\newacro{SA}{ short axis}

\newacro{GPPA}{ generalized partial procrustes analysis}

\newacro{PDM}{ point distribution model}
\newacro{CV}{ cross validation}

\newacro{DR}{ dimensionality reduction}

\newacro{BMI}{body mass index}
\newacro{BSA}{body surface area}

\newacro{LV}{left ventricle}
\newacro{RV}{right ventricle}
\newacro{EF}{ejection fraction}
\newacro{HR}{heart rate}
\newacro{RVOT}{right ventricular outflow tract}
\newacro{HCM}{hypertrophic cardiomyopathies}

\newacro{CO}{cardiac output}

\newcommand{\beginsupplement}{%
        \setcounter{table}{0}
        \renewcommand{\thetable}{S\arabic{table}}%
        \setcounter{figure*}{0}
        \renewcommand{\thefigure}{S\arabic{figure*}}%
     }

\section{Introduction}
Cardiac myocites thicken and/or elongate \citep{Arts1994,Grossman1975, Opie2006} as a reaction to altered stimuli (such as pressure or volume loading). These cellular changes are aggregated at organ level, producing size and shape changes in the heart chambers and affecting cardiac function. This process is commonly referred to as cardiac remodelling. Adaptive remodelling helps the heart to pump enough blood to satisfy the system's oxygen demands while maintaining pressure within physiological range. Maladaptive remodelling results in changes that are either not sufficient to compensate altered situations, or start damaging the heart, ultimately making it unable to satisfy the systemic oxygen demands and leading to failure. 

Even if remodelling \textcolor{black}{is an} adaptation of individual myocytes \textcolor{black}{responding} to \textcolor{black}{local} stimuli, in clinical practice \textcolor{black}{they are} often simplified to global changes, \textcolor{black}{such as total ventricular volume or myocardial mass, used to assess volume and pressure loading respectively}. \textcolor{black}{ These global measurements are enough to quantify the most typical remodelling patterns:} a wall thickening  with inward motion of the inner wall (thus reducing cavity size and wall stress) as a reaction to pressure loading and a dilatation of the cavity to cope with \textcolor{black}{chronic} volume overload, so that, \textcolor{black}{for equal} wall deformation during contraction, more stroke volume is ejected with each beat. However, many conditions have been described to produce regional remodelling: such as the presence of a basal septal bulge induced by hypertension \cite{Baltabaeva2008}, a thickening of the septal wall in hypertrophic cardiomyopathies \citep{Olivotto2012} or a base-to-apex gradient in deformation in deposition diseases such as amyloidosis \citep{Cikes2010} or thalassemia \citep{Hamdy2007}. In the clinical community, the quantification of regional patterns is mostly used in segmental motion and deformation assessment (particularly in coronary artery disease) rather than local geometry. For the \ac{LV}, a standardised partition in 17 wall segments has been proposed \citep{Cerqueira2002}, which recently also has been used to quantitatively assess regional strain patterns. However, these wall segments are by definition equal in size and thus of limited utility to assess changes in morphology. 

In medical image analysis, a common method for quantifying regional morphology is the creation of an \emph{atlas}, which is a template shape representative of a population, and registering each patient-specific shape to this atlas \citep{Zhang2017}. This has been successfully applied to describe inter-individual \textcolor{black}{gross} variations of morphology in populations, but it is more challenging to follow subtle regional remodelling within an individual over time, \textcolor{\colorComments}{since the use of an atlas regularises and smooths the data}. Additionally, this approach has the drawback that it requires registration (deforming the atlas to match the individual) \citep{Joshi2004}. Registration remains an unstable and computationally expensive process, even if there has been  advances to ease both the stability and computation time \citep{Benkarim2019,Fu2019} A particular problem of the cardiac \ac{RV} is that there are few landmarks that can be identified among different individuals: registration is based on image intensity/shape patterns, and has no guarantee that the established point-to-point correspondences \textcolor{black}{have an anatomical meaning}. Thus, when using computational meshes to represent the heart through image segmentation, after atlas registration, an important part of the mesh nodes' positions do not correspond to identifiable anatomic landmarks among different individuals and theirs point-to-point correspondence cannot be used to assess specific physiological remodelling associated with local stimuli. 

To avoid explicit registration, some authors have proposed parametrisation methods to create anatomical maps of organs \citep{Nunez-Garcia2019,Vera2014, Paun2017, Bayer2018, Auzias2013, Hurdal2009} by finding smooth bijective maps from each surface/volume to a common domain, a subset of $\mathbb{R}^2$ or $\mathbb{R}^3$ respectively. The parametrisations of individual anatomies can subsequently be used to obtain a point-to-point correspondence. The mapping is typically obtained through a minimisation of a distortion metric. In particular, several authors propose to force the mapping to be as conformal as possible, ie locally maintaining angles, \citep{Levy2001,Gu2010}. Maintaining angles is not the only possibility: there are other approaches that try to maintain, for instance, local distances \citep{Sorkine2007} or areas. 

Compared to the \ac{LV}, the \ac{RV} has a complex and irregular shape, with more segmental variability \citep{Haddad2008}, making regional analysis of the \ac{RV}  more difficult than of the \ac{LV}. Its position in the chest also produces a bad acoustic window when using 3D echo-cardiography and results in low-quality images, limiting the applicability of complex techniques. Instead, in clinical research, a segmental approach is more common due to its easier interpretability and the spatial smoothing effect of averaging over the segment. There have been studies referring to a regional analysis of different segments that compose the \ac{RV} \citep{Addetia2016, Addetia2018, Moceri2018}. These studies only analyse the endocardium, more specifically its wall curvature. While local curvature is an important component of the wall stress generated by pressure and thus very important in pressure-overload remodelling, an analysis in terms of regional dilatation/volume is needed to correctly assess remodelling originated from volume-overload. \textcolor{black}{Registration techniques are not well suited for volumetric analysis of the \ac{RV} since there are no identifiable patterns in the blood pool.} Therefore, there is a need for quantitative approaches that can assess regional shape and volume changes in a clinically relevant and physiologically plausible way.


In this paper we develop a segmental analysis tool suitable for a clinical application, using images of suboptimal quality, by proposing an automatic method for mesh-independent volumetric parcellation of the \ac{RV} based on the geodesic distance to three easily identifiable anatomical landmarks: tricuspid valve, pulmonary valve and apex. 
Complementarily, in order to assess the method's potential to identify clinically relevant and physiologically plausible regional remodelling, we propose a method to generate synthetic datasets through regional induction of circumferential and longitudinal elongation and analyse the sensitivity of the parcellation to both global and regional remodelling. 
To assess the performance, and robustness to noise, in a real setting, we also do an inter- and intra- observer reproducibility analysis, as well as a test/retest comparison of two sequential acquisitions, on the same patient.  This technique enables both a regional analysis of anatomy, using the \ac{ED} volumes, as well as function, via regional \ac{EF}. 


\section{Methodology}

\subsection{Parcellation of the right ventricle}
\label{subsection:parcellation}
The \ac{RV} has a complex and asymmetric shape \textcolor{black}{that} is positioned partly surrounding the \ac{LV}. Its anatomy is most commonly described as biaxial: one axis goes from tricuspid valve to the apex, and the other from the apex to the pulmonary valve. 
The \ac{RV} can be grossly separated in 3 main anatomically and functionally different parts: the outflow infundibulum, the smooth inlet and the trabeculated apex \citep{Haddad2008, Haddad2008a}. However, there is no consensus on the exact border between these parts, and different experts can draw different region boundaries over the same images.  Given that the partition definition is needed to quantify volume and shape changes over time when doing follow-up in individuals, we propose an automatic method for volumetric parcellation of the \ac{RV}, based only on geometric properties. While this parcellation method has the advantage of being fully automatic and therefore completely reproducible under the same image and segmentation,  the method still relies in the  generated 3D model thus depending on the original segmentation quality and stability. To avoid errors due to a bad point-to-point correspondence, our parcellation does not use explicit registration and only uses the geodesic distances from anatomic landmarks that can be clearly identified in 3D echocardiography: the apex, tricuspid and pulmonary valve. The method is independent of the triangulation of the ventricular surface. 

\begin{figure*}[!tb]

 \includegraphics[width = .99\textwidth]{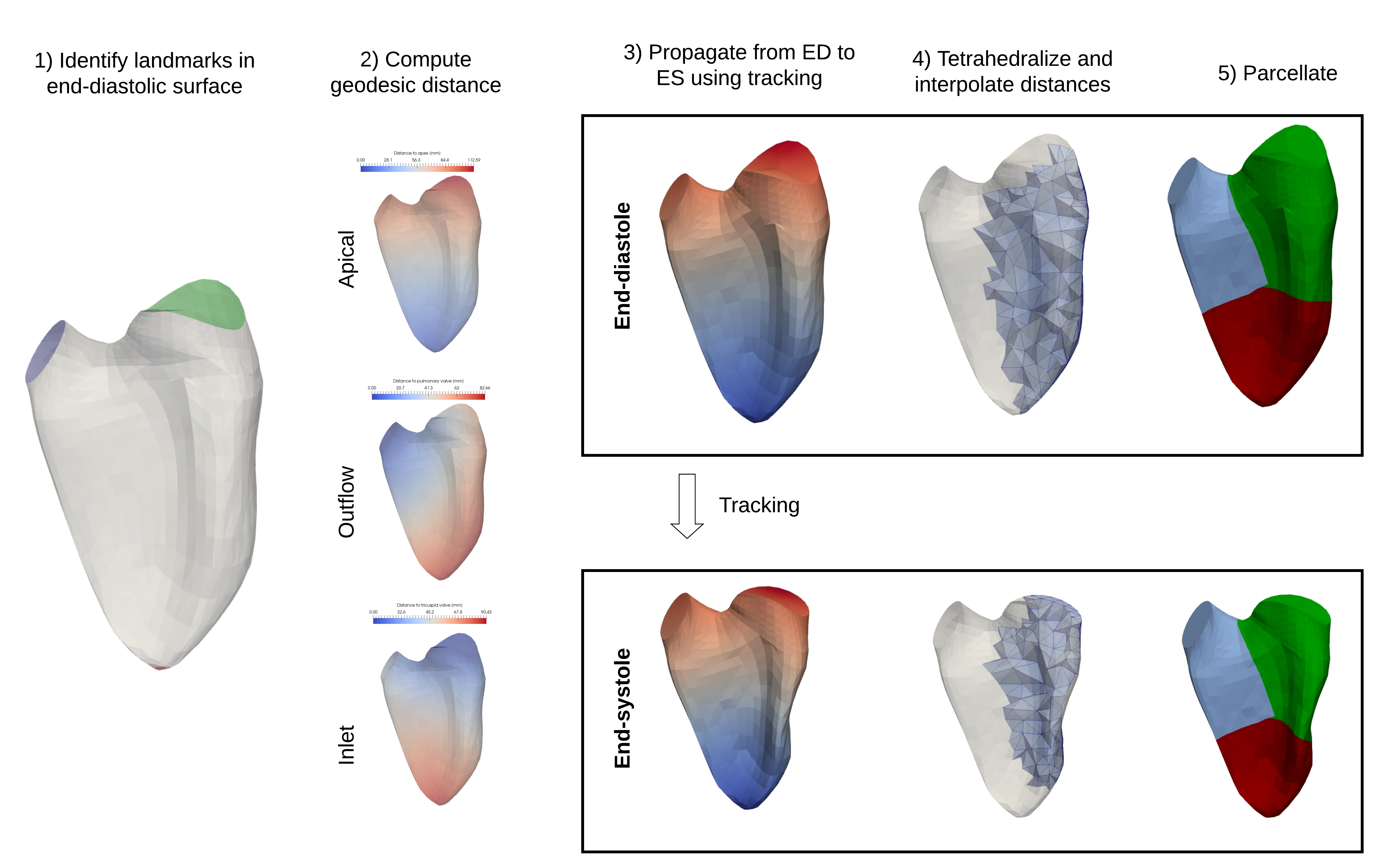}
 \caption[Steps to generate the volumetric partition.]{Steps to generate the volumetric partition. 1-2) For each point of the ED surface,  the geodesic distances to the apex, tricuspid and pulmonary valves are computed as scalar maps. 3) The maps are propagated from the ED surface to the ES surface following the tracking  4) These distance maps are extended from the surface to the cavity by tetrahedralising the ED and ES meshes, and using the  Laplace equation to interpolate values to the interior. 4) The ventricle is split in the regions by assigning each point of the cavity to the closest landmark.}
 \label{figure:rvParcellation:framework}
\end{figure*}

We applied our method to the analysis of 3D models of the \ac{RV} generated from 3D echocardiography images using Tomtec software (4D RV-FUNCTION ), but it can be easily adapted to other processing platforms and imaging modalities. \textcolor{\colorComments}{Tomtec software provides a semi-automatic model-based segmentation of the \ac{RV}, and a tracking during the full cardiac cycle.} Figure \ref{figure:rvParcellation:framework} shows the full process used to parcellate the \ac{RV}, which is described below.

The first step of our parcellation is the identification of the valves and apex using the point-to-point correspondence provided by the semiautomatic segmentation software in the endocardial surface mesh (given that these correspond to stable anatomical landmarks in the mesh that can be identified in the image). Next, for each node of the mesh, we compute the geodesic distances to the apex, pulmonary valve and tricuspid valve. The geodesic distances between two points are computed on the surface with an exact algorithm \citep{Surazhsky2005} that computes the length of the minimal on-surface path between two points. The distance between a point and an anatomical substructure is defined as the minimum distance from the point to any point that belongs to that anatomical substructure:
\begin{equation}
\begin{array}{lll}
    d_{t}(x) & =& min \{ d_{geo}(x, y) | y \in  \textit{tricuspid valve} \} \\
    d_{p}(x) &=& min \{ d_{geo}(x, y) | y \in \textit{pulmonary valve} \} \\
    d_{a}(x) &=& min \{ d_{geo}(x, y) | y \in \textit{apex} \}
\end{array}
\end{equation}

Figure \ref{figure:rvParcellation:framework}.a shows the geodesic path from an arbitrary point to the different landmarks, and Figure \ref{figure:rvParcellation:framework}.b shows distance from every point of the surface to the apex represented as a heatmap: the points furthest to the apex are coloured in red and the closest in blue. After this distance is computed for every point of the surface, the interior of the triangular surface mesh is tetrahedralised using a publicly available software (TetGen version 1.5.1, \cite{Si2015}) (Figure \ref{figure:rvParcellation:framework}.c). The distance defined on the surface of the mesh is propagated to the interior of the ventricle using Laplace's equation.  The equation uses the tetrahedralised mesh as domain and Dirichlet boundary conditions specified by the surface-defined distance maps. Formally, this interpolation step is defined as follows, where $M \in \{apical, inlet, outflow\}$, $\Omega$ refers to the volumetric domain and $\partial \Omega$ and $\mathring{\Omega}$ to its boundary (the surface mesh) and interior respectively:
\begin{equation}
\left\{ \begin{matrix}
 \Delta u_{M} = 0 \;  & for \;x \in \mathring{\Omega} \\
 u_{M}(x) = d_{M}(x)\;  & for \;x \in \partial \Omega
 \end{matrix} \right.
\end{equation}

This equation is discretised using finite elements with a publicly available software (Sfepy \cite{Cimrman2019}). This process is repeated to compute and extend to the interior of the cavity the 3 distances. Once the distances are defined in the volumetric mesh, we partition the ventricle, assigning each point of the mesh to the "closest" landmark, using the interpolated distances. M\textsubscript{inlet}, M\textsubscript{apical} and M\textsubscript{outlet}  respectively, are the partition corresponding to the inlet, apex and outlet. Each point is assigned to the region whose representing landmark is closer as shown in Figure \ref{figure:rvParcellation:framework}.4. The partition does not follow the mesh vertices and edges, but new elements are generated during the partition.  We used linear interpolation to define the distance values inside each tetrahedron. A formal definition of the segments is:

\begin{eqnarray}
    \begin{split}
 M_{inlet}=  \{ x  \vert  d_{t} \left( x \right) \le  d_{p} \left( x \right) , d_{t} \left( x \right)  \le d_{a} \left( x \right)   \}
 \end{split} \\
 ~
\begin{split}
 M_{outlet}=  \{ x  \vert  d_{p} \left( x \right) \le  d_{t} \left( x \right) ,  
 d_{p} \left( x \right)  \le d_{a} \left( x \right) \} 
 \end{split} \\
 ~
 \begin{split}
  M_{apical}=  \{ x  \vert  d_{a} \left( x \right) \le  d_{t} \left( x \right) , 
  d_{a} \left( x \right)  \le d_{p} \left( x \right)  \} \end{split}
\end{eqnarray}
 
This \textcolor{black}{endocardial} partition can be propagated from \textcolor{black}{\ac{ED} to \ac{ES}} using the point-to-point correspondence between surfaces belonging to the same individual, that are obtained via tracking the initial surface, and then extended to the interior cavity via the same Laplacian interpolation. With this procedure, we can compute regional \ac{ES} volumes and ejection fractions, allowing for regional functional assessment of the \ac{RV}.

\subsection{Local and global anatomic frame of reference}
\label{section:anatomicSOC}

To clinically interpret local geometric changes, it is more convenient to \textcolor{black}{use} an anatomical frame of reference (longitudinal and circumferential), instead of the Cartesian system of coordinates. At each point of the mesh, circumferential and longitudinal directions are defined locally using the method proposed by \cite{Doste2019}. We defined the longitudinal direction using the stationary heat flow in surfaces, with a cold source in the apex, and two hot sources at the same temperature in  the two valves. \textcolor{black}{An alternative approach to compute the longitudinal directions, based on the geodesic directions is discussed in Supplementary Material S6. }The heat flow is computed by solving the  Laplace-Beltrami linear differential equation on a surface. The  Laplace-Beltrami operator ($\Delta$) is discretised using the cotangent formulation \citep{Pinkall1993}:

\begin{equation}
\left\{\ \begin{matrix}\ \Delta u = 0 \ \\ u(apex) = 0  \\ u(valves) = 0 \end{matrix}\ \right.
\end{equation}

The longitudinal direction ($\overrightarrow{l}$) at each point is the result of normalising the resulting temperature gradient. The circumferential ($\overrightarrow{c}$) is chosen to be orthogonal to both the longitudinal and the surface normal at that point ($\overrightarrow{n}$), so that ($\overrightarrow{l},\overrightarrow{c}$)) form a base of the tangent space at the given point: 
\begin{eqnarray}
\overrightarrow{l}=\frac{\nabla u}{ \Vert u  \Vert} \\ 
\overrightarrow{c}= \overrightarrow{l}  \times \overrightarrow{n}  
\end{eqnarray}

\begin{figure}
\begin{center}
\begin{subfigure}{0.2\textwidth}	
    \includegraphics[width=0.95\textwidth, trim = 12.5cm 4cm 12.5cm 4cm, clip]{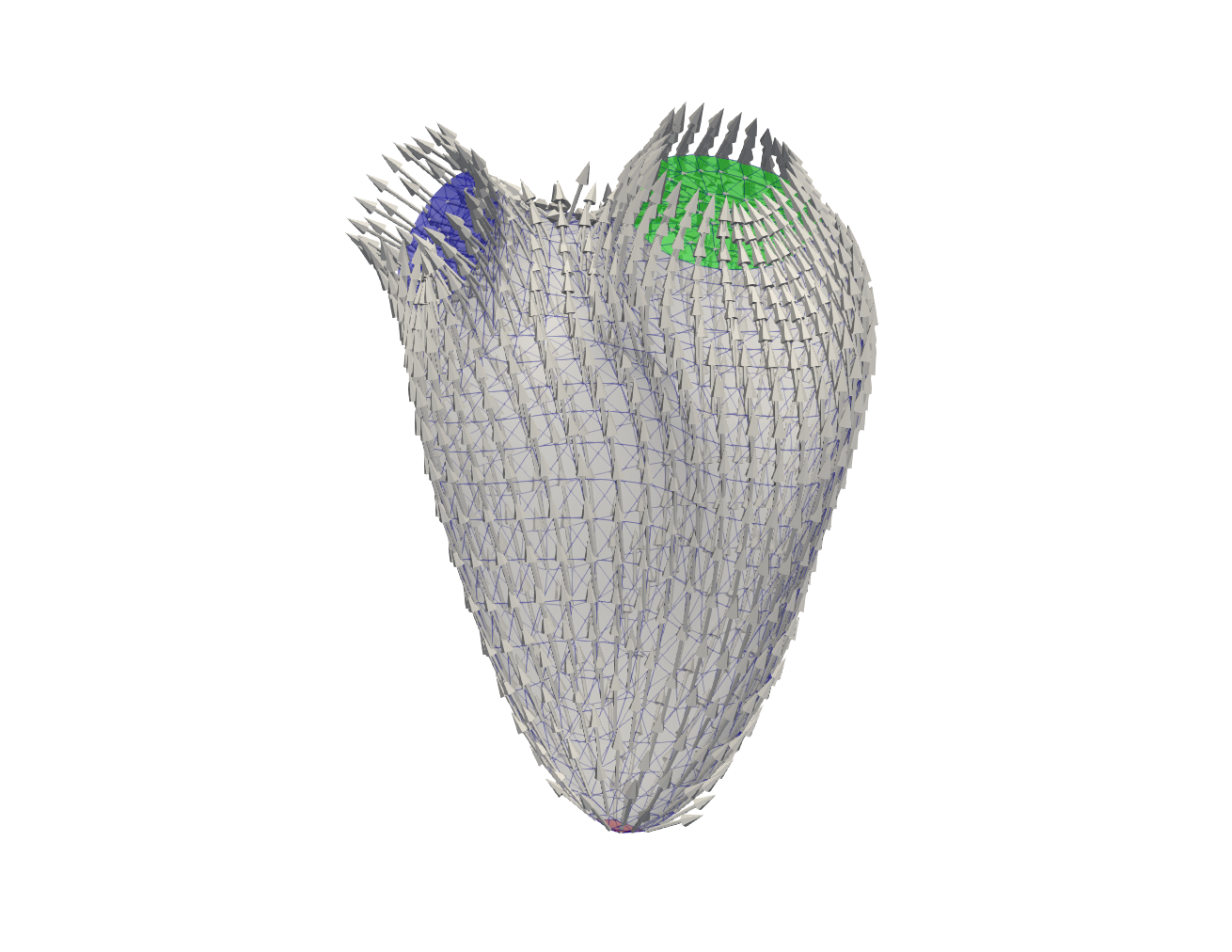}

\end{subfigure}
~
\begin{subfigure}{0.2\textwidth}	\includegraphics[width=0.95\textwidth, trim = 12.5cm 4cm 12.5cm 4cm, clip]{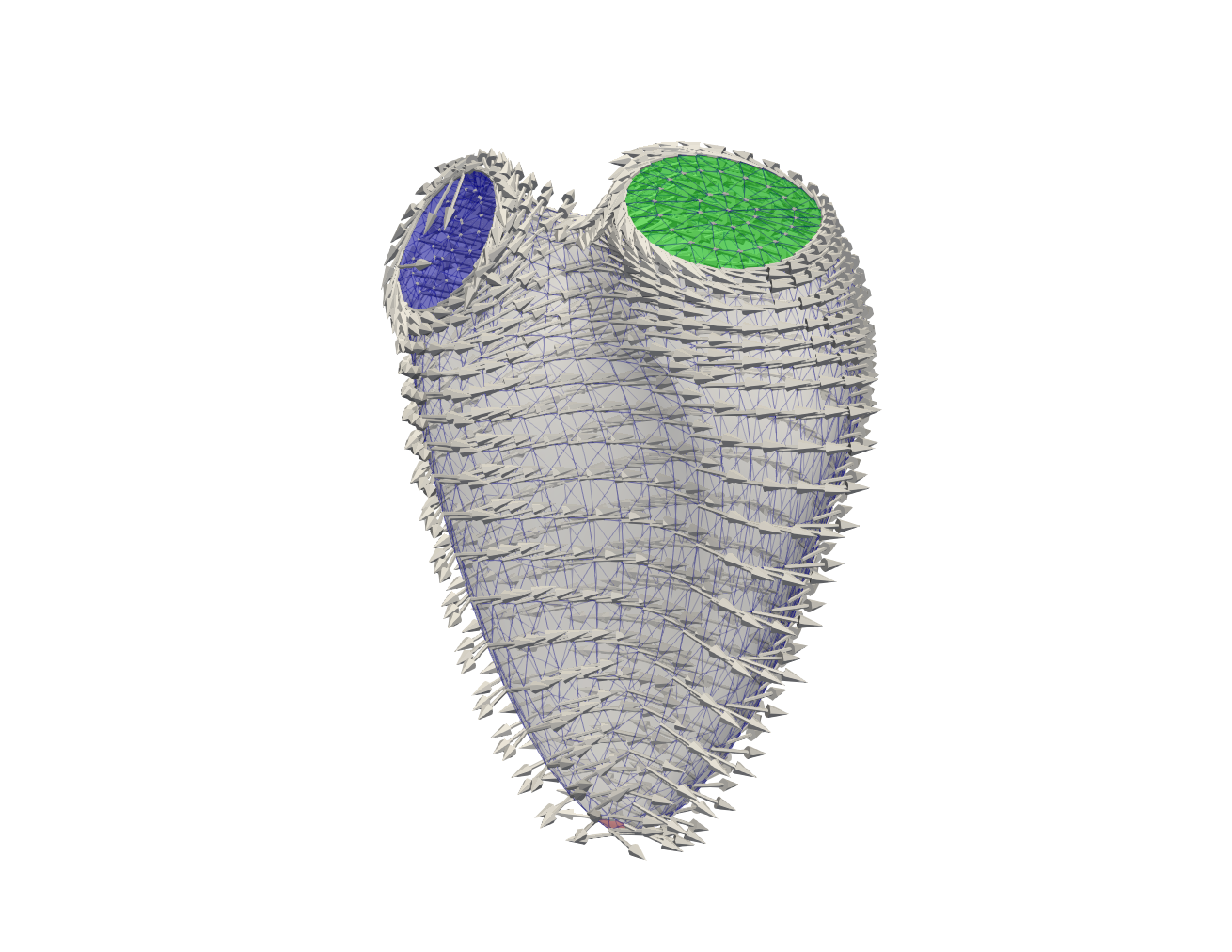}
	\end{subfigure}

\end{center}

~		\caption[Longitudinal  and circumferential directions defined in each triangle of a sample RV mesh]{Longitudinal (\textcolor{black}{left}) and circumferential (\textcolor{black}{right}) directions defined in each triangle of a sample RV mesh. The tricuspid valve is shown in green  and the pulmonary valve in blue. Since the valves are not part of the myocardium, the definition of the anatomical direction there has no meaning. }
    \label{rvMethodological:fig:directions}
\end{figure}

Figure \ref{rvMethodological:fig:directions} shows the local circumferential and longitudinal directions. A global longitudinal direction is computed by averaging all the local ones, and the circumferential directions are defined as orthogonal to the longitudinal ones.


\subsection{Remodelling deformation strain}
Given two meshes, a reference and a remodelled mesh, in point-to-point correspondence, we can compute the strain associated to the remodelling. This strain fully characterises the deformation, modulo rigid body transformations. To interpret this strain, it is more natural to work in the previously defined local anatomical reference frame. For each triangle, we can express each of its edges as a combination of the local anatomical directions (l, c). We note $E_{t}^{m} $  the matrix that contains the edges of triangle $t$ and mesh $m$. With this, we can compute the local linear transformation  $F_t$ at triangle $t$. In a continuous setting, this $F_t$ corresponds to the Jacobian matrix of the deformation.
\begin{equation}
    F_t = E_t^{m_{def}} \left( E_t^{m_{ref}} \right)^{-1}
\end{equation}

From this transformation, we can compute the Cauchy strain tensor for small displacements $\epsilon = (F^t + F)/2$, and extract the  strain in the longitudinal ($\epsilon_{ll}$) and circumferential  ($\epsilon_{cc}$) directions. Note that the longitudinal and circumferential directions are not necessarily aligned to the principal strain directions, which are the eigenvectors of the strain tensor.

\subsection{Generation of synthetic remodelling patterns}
To test our algorithm, we apply a prescribed strain to a real reference mesh, obtaining a synthetic remodelled mesh for which we know which are the areas \textcolor{black}{that deformed}, and with which magnitude. This is achieved by solving an inverse problem of the forward strain calculation defined in the previous section. We used a modification of a linear surface construction method  \citep{Wang1981} that  generates 3D meshes from a local description. The remodelled mesh connectivity is an input of the method and it cannot contain a non-manifold edge. The local description consists of the edge lengths and the dihedral angle associated to every edge (the angle formed by the normals of the adjacent triangles). Obviously, not all combinations of lengths and angles define a valid surface, but we can formulate the reconstruction in a minimisation setting so we obtain the surface satisfying as much as possible the local description. We will use this method to simulate different types of remodelling, by locally deforming a template mesh.

\subsubsection{Local descriptors of the surface}
\label{section:localDescription}
We describe a local frame of reference for each triangle of the mesh from the input data. This frame is arbitrary but uniquely defined, assuming a unique ordering of the nodes inside each triangle. It uses the first node of the triangle as origin, the direction of the first edge as x-axis, its in-triangle normal as the y-axis and the z-axis is the triangle normal. We will call $a_{t,i}$ the coordinates of the i-th point of the t-th triangle expressed in local frame. By convention, $a_{t,0} = (0,0, 0)$, the third coordinate is always 0 for all points (since they are coplanar), and $a_{t,1} = (x, 0,0)$ . Note that by basic trigonometry we can compute the local coordinates of the triangle nodes given the 3 lengths, using the constraints that the first node is in 0, and the second one lies in the x-axis. 

\begin{figure}
	\begin{center}
		\includegraphics[width=.4\textwidth, trim= 14cm 5cm 5.5cm 5cm, clip]{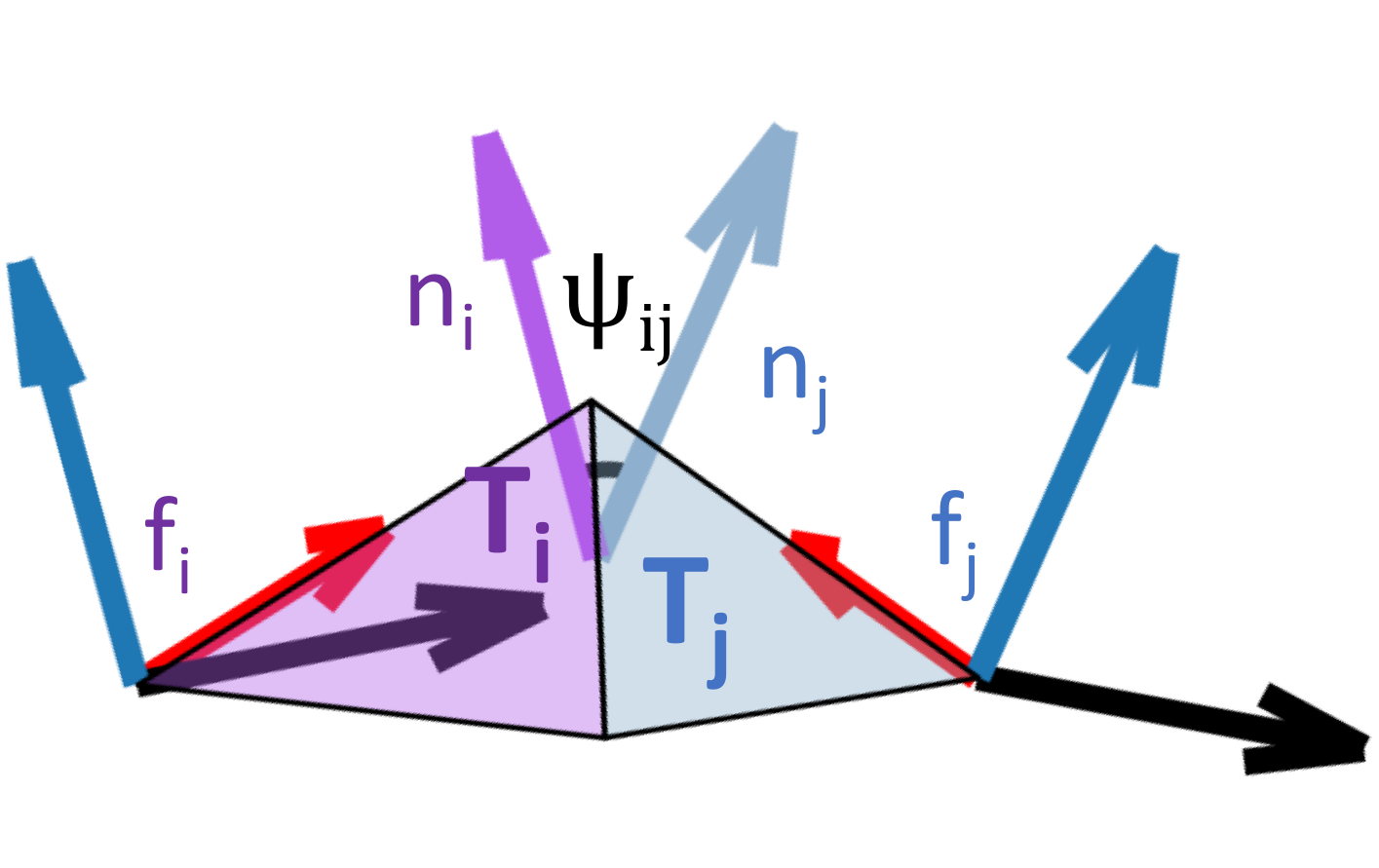}
	\end{center}
\caption[Local frames of reference and dihedral angles.]{Two  adjacent triangles $T_i$ and $T_j$, with their respective local systems of reference, $f_i$ and $f_j$. $\psi_{ij}$ is the dihedral angle, represented at the mean point of the common edge, and it is the angle formed between the two normals ($n_i$, $n_j$. }
\label{figure:meshReconstruction:triangles}
\end{figure}

The unknown variables are the 3D coordinates $x_i$ for each mesh point $i$, and a reference frame $f_t$ associated to each triangle $t$, that corresponds to the mapping from the triangle coordinates $a$ to the 3D space.  Figure \ref{figure:meshReconstruction:triangles} shows two adjacent triangles, with their dihedral angle and the associated frames of reference.

For two adjacent triangles $i$ and $j$, we can obtain the rotation $R_{ij}$ from frame $f_j$ to $f_i$:
\begin{equation}
\label{eq:reconstruction:RDefinition}
  R_{ij} =f_{j}^{T}f_{i} 
\end{equation}

We can express $R_{ij}$  using only the local descriptor: the triangle coordinates and the dihedral angles $\psi_{ij}$. We call $\phi_v^{\theta}$ the rotation of angle $\theta$  around the axis of rotation $v$. Let $e$ be the common edge between triangles $T_i$ and $T_j$, we can compute the angle $\theta$  between edge $e$ and the first edge of $T_i$, and $\theta'$  is the respective angle for $T_j$. Then, we can express $R_{ij}$ as the composition of $3$ rotations: 

\begin{equation}
R_{ij}= \phi_{z}^{ \theta'} \phi_{x}^{\psi_{ij}} \phi_{z}^{ \theta}
\end{equation}

\subsubsection{Linear reconstruction}
In this section we formulate the inverse problem to generate a mesh using the previously defined variables $R$ and $a$. Given a mesh topology, and values for the dihedral angles and edge lengths defined at each edge and face respectively, the algorithm finds a 3D mesh that satisfies as much as possible the desired local geometry. We manipulate the forward equations of the preceding section to obtain an equivalent form which is linear on $x$ and $f$, and use their quadratic residual as energy to minimise, thus formulating an optimisation problem with the world position of the nodes $x$ and the frames $f$ as variables. To easily solve this problem, we do not enforce that matrices $f$ are rotations, but arbitrary matrices. 

By multiplying by $f_j$ equation \ref{eq:reconstruction:RDefinition} and rearranging terms, we obtain the following equivalent equation:
\begin{equation}
\label{rvMethodological:eq:f}
 f_i - f_j R_{ij} = 0
\end{equation}

For every edge  $e_{ij}$ in every triangle  $t$, where the nodes have in-triangle indices i’ and j’ respectively, we can use that $f_t$ to transform from triangle coordinates to world coordinate:
\begin{equation}
 x_{i} - x_j =  f_ t \left( a_{ti'} -  a_{tj'} \right) 
\end{equation}

After moving all terms to the LHS, we obtain:\par
\begin{equation}
\label{rvMethodological:eq:energyLinX}
 x_{i} - x_j -  f_ t \left( a_{ti'} -  a_{tj'} \right)  = 0
\end{equation}

Thus, we have obtained equations for computing $x$ and $f$ from $R$ and $a$. We create an energy by minimising the squared $L_2$ error of the sum over all edges of  equations \ref{rvMethodological:eq:f} and \ref{rvMethodological:eq:energyLinX}. For the matrices we use the Frobenius norm, which is simply the sum of squares of all the elements. We can add weights to each term of the equation (the one that solves for the frames, and the one that solves for the node position) to control their relative contribution to the global solution. This energy is quadratic and sparse, so it can be efficiently solved with linear methods via its normal equations. Its final form reads:

\begin{eqnarray}
\label{rvMethodological:eq:linEnergy}
W_I(x, f) = \sum_{ij \in E} \sum_{\forall t \adjacent ij}   \| x_i - x_j - f_t(a_{ti'} - a_{tj'}) \|^2 \\
W_{II}( f) = \sum_{ij \in E} \| f_i - R_{ij} f_j \|^2 \\
W = \lambda_1 W_I + \lambda_2 W_{II}
\end{eqnarray}

Where $E$ is the set of edges of the mesh, and the predicate $t \adjacent ij$ denotes that the face $t$ is adjacent to  edge $ij$ and $a_{ti'}$ (respectively $a_{tj'}$) denotes the in-triangle coordinates of node $i$ (respectively $j$) in face $t$. \textcolor{\colorComments}{Since the meshes obtained from the segmentation were very regular, we have set both $\lambda_1$ and $\lambda_2$ to 1.}

\subsubsection{Log-domain reconstruction}
In the previous formulation, it is not imposed that frames $f$ exactly correspond to orthonormal matrices. This can lead to a degenerate solution where all $f$ have a determinant $< 1$ and the mesh is shrunk. To avoid this situation, a solution is to enforce that matrices $f$ are rotations by parametrising them on an appropriate domain.


Any scalar analytic function can be converted to a matricial function, by using the matrix multiplication instead of the usual multiplication to compute the powers of the variable in the Taylor series of the function. Therefore, the matrix exponential is defined as:
 \begin{equation}
    exp \left( X \right)  =  \sum _{n}^{}\frac{X^{n}}{n!}
\end{equation}

A well known result is that the matrix exponential of a matrix $A$ is a rotation if and only if $A$ is antisymmetric, so, we can make $f_i = exp(w_i)$, where $w_i$ is an antisymmetric matrix. An antisymmetric matrix can be parametrised by its 3 lower triangular components. We define $[v]_{\times}$ as the mapping from the parameters v to the associated  antisymmetric matrix, where $v \in \mathbb{R}^3$:

\begin{equation}
[v]_{\times} = \begin{pmatrix}
0 & -v_3 & v_2 \\
v_3 & 0 & -v_1 \\
-v_2 & v_1 & 0 
\end{pmatrix}
\end{equation}

After this parametrisation  the previous energy is no longer quadratic (Eq \ref{rvMethodological:eq:linEnergy}), but we still can compute its derivatives. Usually matrix functions are very cumbersome to differentiate, but for the particular case of the matrix exponential of a 3D antisymmetric matrix, there exists a closed-form expression \citep{Gallego2015}. Specifically, when applied to a fixed vector $u$, and using the notation $f(v) = \exp([v]_{\times}$, where $v$ are the reduced parameters of an antisymmetric matrix:

\begin{equation}
\label{eq:logDerivative }
    \frac{\partial f(v) \cdot u}{\partial v} =  -f(v) [u]_{\times} \frac{vv^T + (f(v)^T - Id)[v]_{\times}} {\| v \|^2}
\end{equation}

Using the derivative formula of the matrix exponential, and standard matrix calculus we find the derivatives of the previous expressions. We use the trick that, for any orthonormal base ($e_0, e_1, e_2$) and any matrices $A, B \in \mathbb{R}^{3\times3}$:

\begin{equation}
\langle A, B\rangle_{F} =  \sum_k \langle A e_k, Be_k\rangle
\end{equation}

The first dot product is the Frobenius product between matrices, and the second is the usual dot product between vectors. With this trick, we can compute the derivatives for each rotation defined over a face $f_i$, and each node position $x_i$. We note as $D(v)[u]$ the derivative $ \frac{\partial f(v) u}{\partial v} $  to avoid a cumbersome notation. After some trivial computations and reordering, we obtain:

\begin{eqnarray}
    &f_i = exp(v_i) \\
    & \begin{split}
    \nabla_{f_i} &W_I(v, x) = \\ 2 \sum_k \sum_{ij\in E}  D(f)[e_k](f_{t_i} e_k  - R_{ij} f_{t_j} e_k)
    \end{split}
    \\
    & \begin{split}
    \nabla_{f_i} W_{II}(v, x) = \\ 2  \sum_{e = (u,v) \in t_i}   D(f)[a_u' - a_v']\left(x_u - x_v - f(a_u' - a_v') \right) 
    \end{split} \\
    &\nabla_{x_i} W_I(v, x) = 2  \sum_t \sum_{e = (i,j)  \adjacent i} \langle x_i - x_j - f_t (a_i' - a_j') \rangle \\
    &\nabla_{x_i} W_I(v, x) =  0
\end{eqnarray}
Where $t_i$ and $t_j$ refer to the faces of node $i$ and $j$ respectively, that are adjacent to the edge $ij$. Since we can analytically compute the gradient, we can use a first order optimisation method. We use the L-BFGS  quasi-Newton algorithm \citep{DennisJr.1977,Nocedal1980}.

\subsubsection{Global remodelling}
Global remodelling involves large changes that correspond to overall size rather than the local shape remodelling. It corresponds to the variability due to, for instance, the individual's height and weight. Even though it could be generated using the previous method, a simple affine transformation applied to the reference mesh nodes coordinates is enough to model this remodelling. First, we define the global longitudinal direction $l_{glob}$ as the average of the local longitudinal direction, and the circumferential directions are defined as orthogonal to the longitudinal \textcolor{black}{ones}. The longitudinal and circumferential remodelling transformations are modelled as linear functions  \textcolor{black}{parametrised by} $t$, which is the stretching factor in the desired direction.  The longitudinal transformation is defined as  ($Id + t  \cdot l_{glob} \otimes l_{glob}$), and the circumferential as $\left( Id + \sqrt{t} \cdot  (Id - l_{glob} \otimes l_{glob}) \right)$, where $Id$ is the $3\times3$ identity matrix, $\otimes$ is the Kronnecker product.

\section{Experimental setting}
\subsection{Data acquisition}
3D echocardiographic images of 10 male \textcolor{black}{healthy American football players without cardiovascular disease} were acquired in a modified apical 4 chambers view using an EPIQ7 ultrasound system (Philips Medical Systems, Andover, MA, USA) equipped with a 1 to 5 MHz transthoracic matrix array transducer (X5-1). For each individual, 4-6 different ECG-gated subvolumes were acquired in a single breath-hold to be compounded into the full 3D+t images of 2 complete cardiac cycles. A written informed consent form was obtained from all study participants.  \textcolor{black}{The demographics of the population can be seen in the supplementary material S1}.

Another male subject\textcolor{black}{, a pure control,} was imaged in two consecutive acquisitions by two different operators to obtain an estimate of the variability due to the imaging process.

The image loops  were processed  using a clinically validated software (4D RV-FUNCTION Tomtec-Arena TTA2, Tomtec Imaging Systems GmbH, Unterschleissheim, Germany \citep{Niemann2007,Muraru2016}) to segment and track the \ac{RV} and obtain a 3D-model for each patient. These models were exportable in \emph{ucd}, a standard file format. All 3D models had the same topology: a triangular watertight mesh with 938 nodes and 1872 faces. The points were in approximate point-to-point correspondence among different individuals. The segmentation pipeline consisted of the following steps:  first the clinician segments the \ac{RV} endocardium contour of the frame corresponding to the R peak using the semiautomatic tool, and the result is tracked during a full cardiac cycle. Afterwards, the clinician can adjust the \ac{ES} and \ac{ED} (defined as the peak of the R wave in the ECG) segmentation iteratively until visually satisfied with the resulting contours.

\subsection{Registration and harmonic maps}
\textcolor{black}{
We additionally assessed  an alternative to parcellate the \ac{RV} based on registration and harmonic maps. For the registration approach, we defined the partition in a template mesh, and used a state-of-the-art software (Deformetrica v4.3 \cite{Bone2018}) to map a parcellation defined in a template mesh to the surface of the target mesh. Afterwards, the steps of tetrahedralisation and interpolation of the partition to the blood pool are computed as described in \ref{subsection:parcellation}. Previous to the non-rigid registration, we aligned the template and the target meshes using the apex and valves positions. 
For harmonic mapping approach, we used the heat maps with sources in each of the 3 landmarks instead of the geodesic distance. The results of the harmonic mapping experiments are found in supplementary material (S6), and were found to be worse than the geodesic approach.}

\subsection{Reproducibility of the semiautomatic segmentation}
We analysed the reproducibility of the \ac{RV} segmentations of the 3D-echocardiographic images processed using Tomtec. For each subject, the same image loops were requantified 3 times for each acquisition by two different operators, to assess the inter- and intra-observer reproducibility of the resulting surfaces and their point-to-point correspondence. For each pair of re-analysed surfaces derived from the same images, we computed the difference of global volumes and the Dice coefficient of the different segmentations, a standard measure for comparing two different segmentations (noted as $S_1$ and $S_2$) and defined as:
\begin{equation}
    Dice(S_1, S_2) = \frac{2\lvert S_1 \cap S_2 \rvert}{\lvert S_1 \rvert + \lvert S_2\rvert}
\end{equation}

As a measure of the mesh node stability, we computed for each node the registration error as the euclidean distance between corresponding nodes from different quantifications as well as the point-to-surface distance (distance from a point to the closest point on the other mesh, that might lie inside a face) for every node. For visualisation, we show the mean point-to-point (resp point-to-surface) error map, averaged over the different individuals. This map was plotted over a template mesh, which was the population mean shape computed via Generalised Partial Procrustes Analysis \citep{Dryden1998}.

Since the test/retest segmentations come from different images, \textcolor{black}{we rigidly aligned the test-retest meshes as explained in the registration subsection, and computed the same metrics used in the inter/intra-observer reproducibility analysis.}

\subsection{Reproducibility of the parcellation method}
The \ac{RV} parcellation computed by our method is dependent on the \ac{RV} segmentation, whose variability was assessed in the previous section. Therefore, even if our method is automatic, we need to evaluate the robustness of the method to the segmentation variability that is present in a normal clinical setting.  

We use the re-requantified dataset to test the inter- and intra- observer reproducibility of the regional volumes and \ac{EF}s. We report the mean and percent absolute error in volumes and \ac{EF}. \textcolor{black}{To assess the stability of the parcellations themselves, we compute the stability of the intersection point of the three segments.}  \textcolor{black}{We compared the reproducibility of our original approach, in which the parcellation was independently computed for each mesh, and a registration-based approach in which the parcellation is first defined in a template mesh, and then transported to each individual using registration.} \textcolor{black}{We also assessed the method's stability to changes in the meshing process: we remeshed the surfaces, and  compared the regional volumes before and after that process. }

We report the regional volumes for both acquisitions obtained in the test-retest acquisition, as well as the absolute and percent differences. 

\subsection{Validation of the parcellation method}
To validate our method, given the lack of a clinical ground truth, we used the deformation method to generate a synthetic dataset \textcolor{black}{including local and global remodelling}. Since we imposed the remodelling, we know the specific areas as well as the exact amount of remodelling, computed as the global difference in volume between the template and the remodelled meshes.

For each \ac{RV} part (apical, inlet, RVOT), we generated two local remodelling patterns: one elongating in the longitudinal direction, and the other one affecting the circumferential direction. These \textcolor{black}{remodelled meshes} were generated using the log-domain reconstruction method. As explained in the methodology section, this method needs as input the mesh topology, the rotation between faces $R_{ij}$ (derived from the dihedral angles) and the in-triangle coordinates for each node $a_{ti}$. The topology and \textcolor{black}{dihedral angles are the same as the ones of the} reference mesh. We define the equation governing the remodelled coordinates $a_{ti}$ in terms of the  in-triangle coordinates of the reference mesh $a_{ti}^{ref}$ and the strain $\epsilon_t(\lambda)$, which will be defined in the upcoming paragraph
\begin{equation}
 a_{ti} - a_{tj} = \left(Id + \epsilon_t(\lambda)\right)( a_{ti}^{ref} -  a_{tj}^{ref})
\end{equation}

The strain $\epsilon_t$ is localised to a segment by imposing a decay on the desired strain magnitude \textcolor{black}{proportional} to a Gaussian function of the distance to the anatomical landmark defining the segment (apex, tricuspid, pulmonary valve). The strain tensor at each triangle is defined as follows:
\begin{equation}
    \epsilon_{t}(\lambda) = \lambda\cdot (v_t \otimes v_t) \cdot \exp{\frac{d_M(x)^2}{\omega^2}} \cdot \exp{\frac{d_{valves}(x)^2}{\omega_{valve}^2}}
\end{equation}

Where $v_t$ corresponds to either the longitudinal or circumferential direction (defined in section \ref{section:anatomicSOC}) at triangle $t$, $\lambda$ is a parameter that controls the remodelling amount and is chosen to satisfy \textcolor{black}{an increment of 5ml} via the regula falsi method. The bandwidth $\omega$ is chosen to be $15$mm.\textcolor{black}{The second factor serves to smoothly impose that the valves do not deform, $d_{valves} = min(d_t, d_p)$ and the bandwidth $\omega_{valve}$ is  set to $7.5$ mm.} The valves' annuli are composed of fibrous tissue and do not show much remodelling in most cases (though they can passively stretch in severe volume overload). Since this localised model is primarily aimed to assess short-term remodelling only, the strain is set at $0$ in the triangles corresponding to a valve.  

The global remodellings were generated by applying the linear transformation corresponding to a total volume increase \textcolor{black}{of $10\%$.}

\textcolor{black}{
We compute the regional volume changes (in ml) from the reference to the synthetic mesh, and compare them to the theoretical ones. For the local remodelling cases, we expect that only the part that was deformed will increase its volume, while in the global ones, all three parts should scale homogeneously. We compute an index of the accuracy as follows:}
\begin{equation}
    acc = 1 - \frac{\sum_{k \in \{apex, base, rvot\} }|(\Delta vols^{computed}_k  - \Delta vols^{theoretical}_k|}{2 \Delta vols_{total}}
\end{equation}
\textcolor{black}{
Where $\Delta vols^{computed}$ refers to the regional volume increment between the remodelled and the original mesh obtained using the parcellation method, and $\Delta vols^{theoretical}$ corresponds to the expected volume increment based on the generation of the synthetic mesh.
This experiment was repeated for both the geodesic-based and the registration-based parcellation approaches. Since in local remodellings there is a big part of the mesh that is not deformed at all, it is easy for the registration algorithm to find correspondences there. This can produce an overestimation of the registration quality. To simulate a more realistic environment, in which the remodellled and original surfaces are meshed independently, we applied a remeshing algorithm \citep{Valette2008GenericDiagrams}. For a fair comparison, the same meshes were used for both geodesic and registration approaches.
}
\textcolor{black}{
To evaluate the correctness of our mesh generation method, we computed the strain from the synthetic remodelled meshes with respect to its reference/baseline. This was compared to the desired strain we imposed, and we report the mean and relative error to the maximal imposed strain.
}

\section{Results}

\subsection{Regional and global volumes of our population}
\textcolor{black}{We computed the regional volumes of the full population. In the suplementary material, we report the populational values in Table S1, and all generated 3D models with their associated parcellations in Figure S1.} There is high variability in both global shape and volumes, especially in the basal part of the inlet and outflow tract.  Visually, the apical segment presents less variability, but the basal region \textcolor{black}{comprised} between the inlet and outflow valves presents more heterogeneity. 

\subsection{Reproducibility of the 3D models}
In this subsection we present the analysis of the inter- and intra-observer reproducibility of the \ac{ED} shapes, as acquired from 3D echocardiography, as well as a qualitative discussion on the test/retest. Table \ref{table:parcelationMethodological:nodeStability} shows the mean error of the different metrics to assess shape differences: total volume difference, Dice coefficient and mean point-to-point and point-to-surface distances. As expected, \textcolor{black}{intraobserver} reproducibility was lower than \textcolor{black}{interobserver}. The total volume error was below $10\%$ for both inter- and intra-observer, but the other metrics, that evaluated local differences, indicated lower stability. In particular, the interobserver Dice coefficient \textcolor{black}{(0.84)} was  lower than the intraobserver one (0.89).

\begin{table}[]
\caption[Intra and inter-observer variability of the segmentations and  node positions.]{Intra-observer, inter-observer and test-retest (a single sample) variability  (in ml and percent) of the segmentations and  node positions.}
\label{table:parcelationMethodological:nodeStability}
\centering
\begin{tabular}{l|ccc}
\toprule
& Intraobs.  & Interobs.   & Test-retest\\
\midrule
Vol. diff. [ml]      & \textcolor{black}{9.0 (5.6\%)}          & 10.9 (7\%)          &  0.31      \\
Dice coefficient             & 0.84          & 0.89   & 0.91    \\
Node-to-node [mm] &  \textcolor{black}{5.5 $\pm$ 1.1}  &   7.2 $\pm$ 1.2 &4. \\
Node-to-surf.[mm] & \textcolor{black}{2.3 $\pm$ .8}   &  2.9 $\pm$ .8 &  1. \\
\bottomrule
\end{tabular}
\end{table}

\begin{figure}[!tb]
    \centering
    \begin{subfigure}{0.45\textwidth}	
        \centering
        \includegraphics[width=0.95\textwidth]{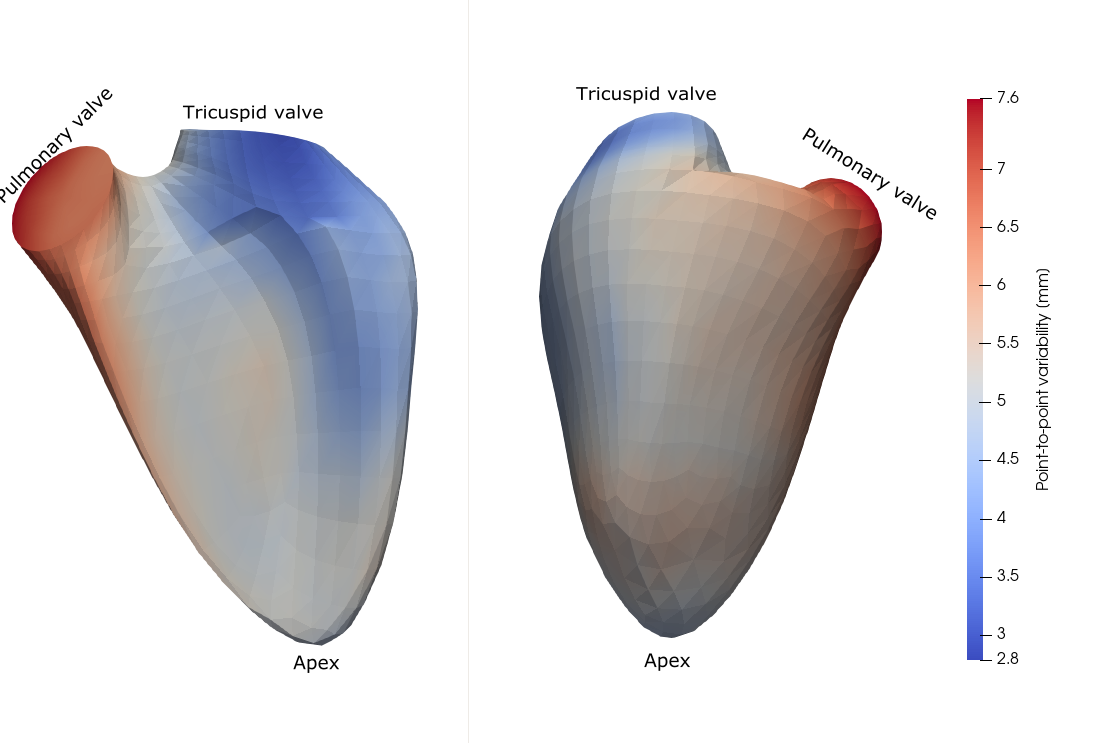}
        \caption[Intraobserver reproducibility]{Intraobserver point-to-point reproducibility}
    \end{subfigure}
    \begin{subfigure}{0.45\textwidth}
        \centering
        \includegraphics[width=0.95\textwidth]{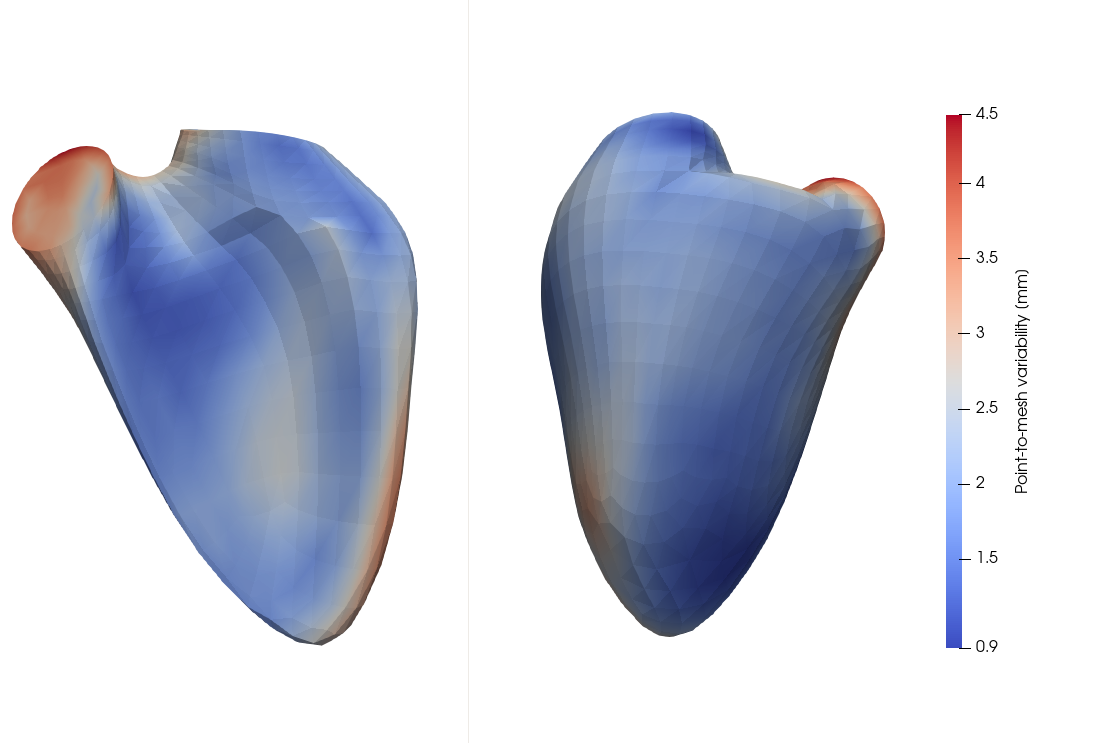}
        \caption[]{Intraobserver point-to-mesh reproducibility}
    \end{subfigure}

    \caption{Mean point-to-point difference for each node on the intraobserver reproducibility test (above), and point-to-mesh distance (below). We can see that the biggest errors are concentrated near the outflow tract, and in the anterior wall. The posterior wall and apical regions are more stable.}
    \label{fig:parcelationMethodological:intraStability}
\end{figure}

\textcolor{black}{Figure \ref{fig:parcelationMethodological:intraStability} shows the regional mean intra-observer point-to-point and point-to-mesh distance obtained in the reproducibility test. We can see that the anterior insertion points, specially near the \ac{RVOT},  present a higher level of instability with mean distances above $1$cm. Instability is not only present in the anterior wall, but also affects the septum. The point-to-mesh error is less spread and more concentrated in the \ac{RVOT}. Figure S2 in the supplementary material shows the inter-observer error distribution, that follows similar patterns to the intra-observer error with higher magnitude.}

\begin{figure}[!tb]
	\begin{center}
		\includegraphics[width=0.45\textwidth]{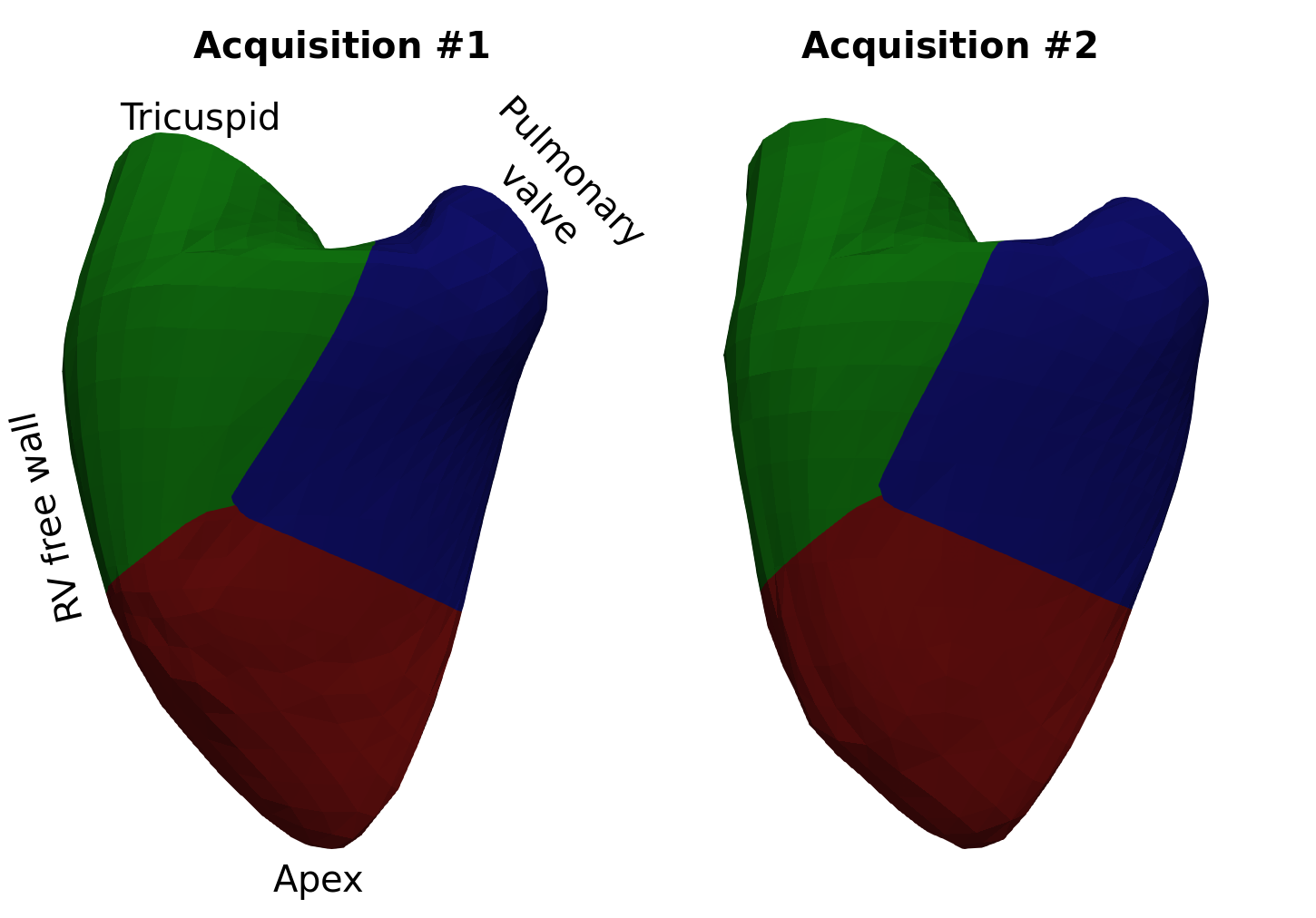}
		\includegraphics[width=0.45\textwidth]{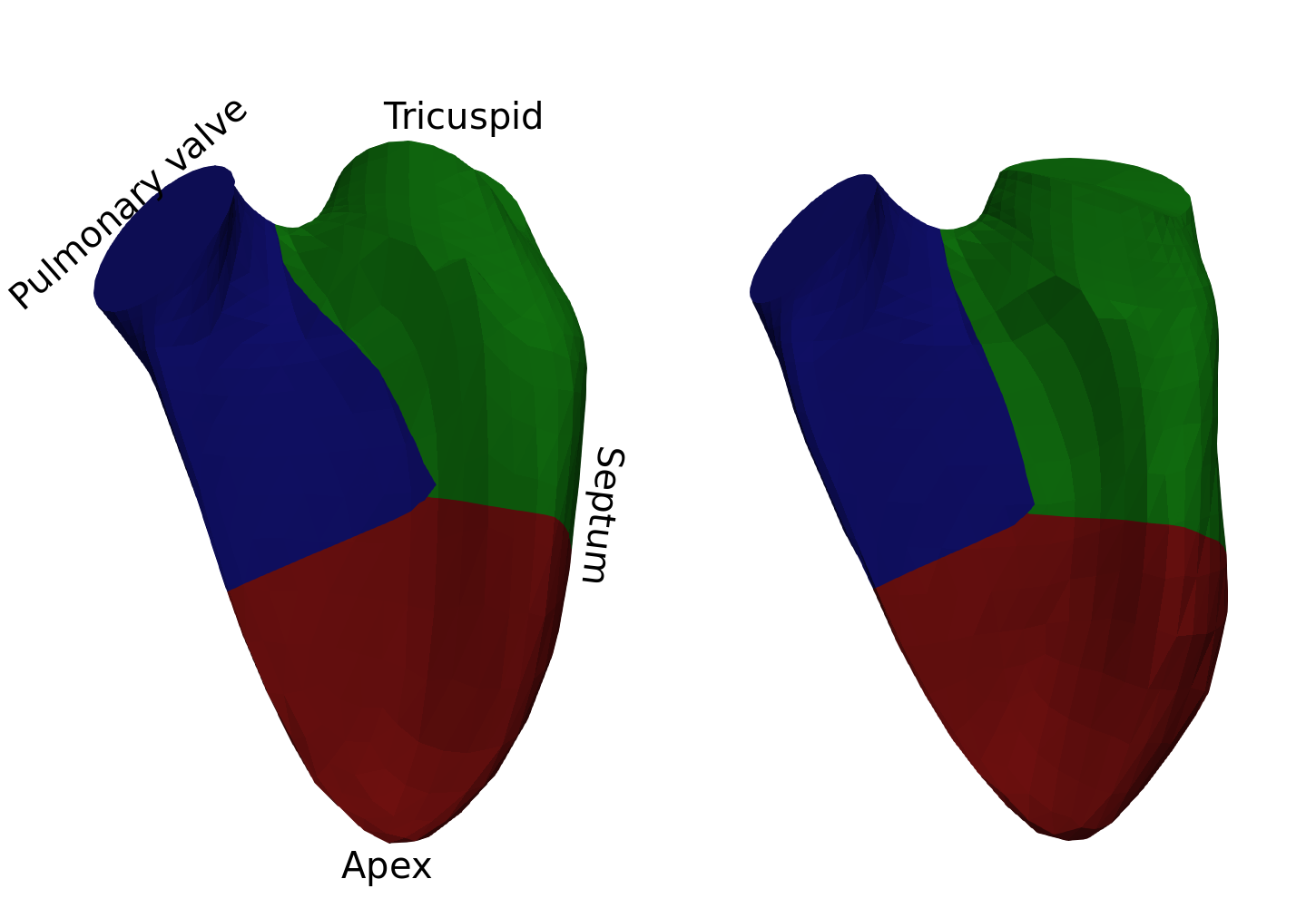}

	\end{center}
\caption[The two generated 3D models on the test/retest experiment and their parcellations. ]{The two generated 3D models  and their parcellations. Even if they have the same approximate size, they have significant differences especially in the anterior wall. These regional differences in the 3D models affects the parcellations: the biggest parcellation differences are in the center of the septum and free wall.}
\label{figure:rvMethodological:testRetestPartition}
\end{figure}

Figure \ref{figure:rvMethodological:testRetestPartition} shows the 3D models generated from the test/retest experiment and their parcellations. \textcolor{black}{Differences in the parcellations will be described in next subsection.} Both meshes have the same total volume, but differ in shape: acquisition \#2 has a higher tricuspid valve and in acquisition \#1 the pulmonary valve is further. Also, acquisition \#1 presents a displaced anterior insertion "line" . Acquisition \#2 segmentation has a wider septum, that extends further in the  anterior segment. Acquisition \#2 has a more spherical apex and a flatter posterior wall. \textcolor{black}{In Table \ref{table:parcelationMethodological:nodeStability} we can find the quantitative metrics of the differences between both acquisitions.  }

\subsection{Reproducibility of the parcellation method}
\textcolor{black}{In this section we study the stability of the regional volumes and \ac{EF}s. Reproducibility was estimated using the inter- and intra-observer reproducibility test dataset}. Table \ref{table:rvMethodological:reproducibilityEDVEF} shows the mean intra- and inter-observer errors and the mean value of the variables for the regional and global \ac{ED} volumes and \ac{EF}. We can observe large errors in the interobserver case ($>$12\%), but  lower for the intraobserver case ($~7-12\%$). The outlet segment had the biggest error for volume(12\%), and regarding function, the apical \ac{EF} \textcolor{black}{had} the biggest error. \textcolor{black}{Both registration and geodesic approaches present similar reproducibility in the apex and inlet segments. Registration was found to be more reproducible in the \ac{RVOT}. We tested the stability of both methods to remeshing: Table \ref{table:rvMethodological:reproducibilityVolumeRemeshing} shows the regional volumes difference between the original and remeshed \ac{RV} surface: registration presented less robustness. }

\begin{table*}[]

\caption[Intra- and inter-observer variability of the segmental and total end-diastolic volumes and EF]{Intra- and inter-observer variability of the segmental and total end-diastolic volumes and EF, expressed as the mean error and mean percent error (in parenthesis).}
\label{table:rvMethodological:reproducibilityEDVEF}
\centering
\begin{tabular}{l|cc|cc|c}
\toprule
 & \multicolumn{2}{c}{Intraobserver error}  & \multicolumn{2}{c}{Interobserver error}  & Mean value \\
{} & Geodesics & Registration  & Geodesics & Registration & \\
\midrule
RV EDV [ml]               &\multicolumn{2}{c|} {\textcolor{black}{9.0 (5.6\%)}}   & \multicolumn{2}{|c|}{10.9 (7.0\%)} &144 \\
RVOT EDV[ml]           & \textcolor{black}{4.7 (12.8\%)} & \textcolor{black}{3.8 (10.2\%)}    & 8.9 (21.9\%)   & 5.0 (12.1\%) &36 \\
Inlet EDV [ml]            & \textcolor{black}{6.4 (7.8\%)} & \textcolor{black}{6.3 (7.6\%)}   & 9.2 (12.2\%) & 7.9 (10.2\%) &68 \\
Apical EDV [ml]           & \textcolor{black}{2.6 (7.0\%)} & \textcolor{black}{2.6 (7.2\%)}     & 3.5 (9.1\%)    & 4.0(11.0\%)   &39 \\
\midrule
Mid-point [mm] & 6.6&  5.5 &8.4 & 9.5 \\
\midrule
RV EF [\%]                & \multicolumn{2}{c|}{3.0 (6.0\%)}  & \multicolumn{2}{|c|}{5.2 (10.2\%)}  &50 \\
RVOT EF [\%]           & \textcolor{black}{3.3 (8.1\%)} & \textcolor{black}{3.1 (7.4\%)} & 11.0 (24.1\%) & 10.9(23.4\%) &42 \\
Inlet EF [\%]             & \textcolor{black}{3.6 (7.6\%)} & \textcolor{black}{3.9 (8.3\%)} & 5.3 (11.2) & 5.3(10.9\%) &61 \\
Apical  EF [\%]           & \textcolor{black}{5.5 (9.7\%)} & \textcolor{black}{5.6 (9.8\%)} & 8.6 (14.8\%) & 8.5(14.4\%) &50 \\
\bottomrule
\end{tabular}
\end{table*}

\begin{table}[]

\caption{\textcolor{black}{Mean absolute differences between the regional volumes obtained by the registration and the geodesic approaches, on the surfaces before and after remeshing, averaged over the full population}}

\label{table:rvMethodological:reproducibilityVolumeRemeshing}
\centering
\textcolor{black}{
\begin{tabular}{c|cc}
\toprule
     & Geodesics & Registration\\
\midrule
RVOT &  0.3 (0.8\%)&   0.7 (1.9\%) \\
Inlet  & 0.55 (0.8\%) & 1.28 (1.9\%)\\
Apical & 0.36 (0.9\%) (& 2.04 (5.2\%) \\
\bottomrule
\end{tabular}
}
\end{table}

We used the test/retest acquisition to verify whether the level of noise is higher in that situation. The two generated 3D models can be found in Figure \ref{figure:rvMethodological:testRetestPartition}, and we can see clear differences in shape. Table \ref{table:rvMethodological:errorDifferenceRecomputed} shows the quantitative analysis of the regional volume differences. The errors found in the test/retest experiment corresponded to the ones observed in the intraobserver reproducibility test.
\begin{table}
\caption[Regional volumes resulting from two consecutive acquisitions of the same patient.]{Regional volumes resulting from two consecutive acquisitions of the same patient.}
\label{table:rvMethodological:errorDifferenceRecomputed}
 			\centering
\begin{tabular}{l|ccc}
\toprule
{} & RVOT (ml) & Inlet (ml) & Apex (ml) \\
\midrule 
Acquisition \#1 & 20.81 & 50.82 & 18.64 \\
Acquisition \#2 & 22.45 & 47.51 & 20.0 \\
Absolute error & 1.64 & 3.25 & 1.40 \\
Relative error & 7.5\% & 6.8\% & 7.3\% \\ 
\hline
\end{tabular}
 \end{table}

\subsection{Generation of synthetic remodelling}
\textcolor{black}{We used the synthetic remodelling method to generate different types of localised (apex, inlet and \ac{RVOT}) and global remodelling, all four cases including pure longitudinal and circumferential dilatations, as well as a homogeneous global scaling. For each \ac{RV} in the dataset, we generated a total of 9 synthetic remodelled meshes (3 global and 6 regional). To assess the reconstruction quality of the local remodelling (the global ones were obtained by applying a linear transformation and therefore their reconstruction is exact), we computed the strain between the synthetic and original mesh, and compared it to the imposed. Table \ref{table:syntheticGenerationError} shows the obtained errors. We observed that the error is below 3\% of the maximal imposed strain in the circumferential cases, and less than 1.6\% in the longitudinal cases. This error was expected, since the imposed strain maps do not allow an exact reconstruction. Examples of the generated meshes, as well as further experiments showing the dependence of the error to the smoothness of the imposed strain field can be found in the supplementary material S3 . }

\begin{table}[]
\caption {\textcolor{black}{Evaluation of the synthetic remodelling generation method, assessed via the mean absolute difference between the imposed and obtained strains, and the relative error to the maximal imposed strain. The maximal imposed strain is used as reference}}
\label{table:syntheticGenerationError}
\centering
\textcolor{black}{
\begin{tabular}{l|ccc}
\toprule
{}  & Mean error  & Rel. error [\%] & Max. strain \\
\midrule 
Apex circ. & 0.42 & 1.05 & 40.04 \\   
Apex long. & 0.38 & 0.49 & 78.71 \\   
RVOT circ. & 0.29 & 2.64 & 11.03 \\   
RVOT long. & 0.29 & 1.42 & 20.28 \\   
Inlet circ. &0.31 & 2.94 & 10.66 \\   
Inlet long. &0.18 & 1.66 & 10.9  \\     
\bottomrule
\end{tabular}
}
\end{table}

\subsection{Validation of the parcellation method}
\textcolor{black}{For all the synthetically remodelled meshes, we computed our parcellation (using the registration and geodesic approaches), and calculated the volume difference for each of its parts with the unremodelled mesh. In Table \ref{rvParcellation:table:validationSyntheticRemodelling} we report the accuracy coefficient of the parcellation to assign the volume increase to the correct segment. The registration template was constructed via applying the Procrustes Algorithm to the remaining meshes. We introduced the remeshing step to avoid overfitting of the registration, the supplementary material S4 contains the equivalent results obtained without remeshing.}

\textcolor{black}{
Results show that the method is mostly accurate for assessment of localised circumferential growth, as well as global dilations. As observed in the reproducibility analysis, the \ac{RVOT} was found to be more noisy and difficult to assess. Comparing registration and geodesic based parcellation methods, we found that registration was more stable at longitudinal remodelling, and also slightly better in the \ac{RVOT}, while the geodesic method was better for circumferential remodelling, specially in the apex.
}

\begin{table}[]
\caption{\textcolor{black}{Accuracy of the parcellation methods (registration and geodesic-based) for identifying the regional volume increase in the synthetic dataset.}}
\label{rvParcellation:table:validationSyntheticRemodelling}
\centering
\textcolor{black}{
\begin{tabular}{l|rr}
\toprule
                    &    Geodesic                & Registration                        \\
\midrule
Apex circ.     &                         \textbf{83.1} &                   49.7 \\
Apex long.     &                             -69.9 &                    \textbf{5.4} \\
RVOT circ.     &              61.4 &                   \textbf{64.5} \\
RVOT long.     &                 -4.2. &                   \textbf{52.2} \\
Inlet circ.     &                   \textbf{85.0} &                   67.2 \\
Inlet long.     &                    42.4 &                   \textbf{65.7} \\
Global circ.   &                  87.5 &                   \textbf{92.1} \\
Global long.   &                    80.9 &                   \textbf{95.8} \\
Global scaling &        \textbf{99.8} &                   95.8 \\
\bottomrule
\end{tabular}
}
\end{table}

\section{Discussion} 

In this paper we proposed a mesh-independent method to volumetrically parcellate the \ac{RV} in three clinically relevant regions: inlet, outlet and apex, with the aim \textcolor{black}{of quantifying} inter- or intra-individual remodelling in regional morphology.

For validation, we additionally presented a method to synthetically remodel meshes in a localised and global manner \textcolor{black}{by specifying a desired strain field}, thus generating a synthetic dataset resembling closely monitoring clinical remodelling. \textcolor{black}{In order for a remodelled mesh associated to a strain field to exist, the strain field needs to verify some conditions (Coddazzi-Mainardi equations, \cite{Wang1981}). Since it is not easy to generate strain fields that satisfy these conditions we generated arbitrary fields, and quantified the reconstruction error and found it below 3\% for the circumferential and 1.6\% for the longitudinal remodelling. Visual inspection of the remodelled meshes showed no artefacts, and a good correlation between the imposed and obtained strains}. From these, we found that the parcellation had a good sensitivity to circumferential or global remodelling, able to attribute it to the correct segment (85\% of the volume was assigned to the correct segment). However, our method is less accurate when analysing local longitudinal localised remodelling. This is not a major problem given that in many clinical scenarios, regional volume and shape remodelling is often in the circumferential direction \citep{DAscenzi2016}.

We studied the variability of the semiautomatic \ac{RV} segmentations. The interobserver reproducibility test of meshes resulting from the segmentation of 3D echocardiography showed that, even if  the mean volume difference was under $10\%$, \textcolor{black} {the Dice coefficient was lower ($0.84$)}. On the other hand, the intraobserver reproducibility  on the same image was higher (Dice coefficient $=$ 0.89). Moreover, the point-to-point (5.5mm) error was much higher than the point-to-surface (2.3mm): the interior nodes of the mesh do not correspond to any anatomical landmark but are equally distributed, thus their correct position does not depend only on the correct segmentation of the \ac{RV} contour in their position, but on the whole segmentation. 

This \ac{RV} variability influences the resulting parcellation, since it is derived from the generated 3D model. Therefore, interobserver reproducibility of the regional volumes and \ac{EF} also \textcolor{black}{presented} a high error ($>$10\%). On the other hand, the  intraobserver reproducibility presented less error: only the outflow volume had a variability $\ge$10\%, while inlet and apex \textcolor{black}{volumes' variabilities} were  7\% . The outflow is complicated to segment given the images often have lower quality there. A qualitative analysis of the partitions showed big differences in the middle of the \ac{RV} since this is the furthest from all landmarks and the method has no information to make the exact parcellation. The inclusion of anatomical landmarks in that area could improve the reproducibility of the method. Regarding the \ac{EF}, the most unstable was the apical part. This is likely  caused by the presence of trabeculations introducing variability in the segmentation as well as the near field effect playing are role there and making the full visualisation of the apex challenging.

The test/retest experiment presented \textcolor{black}{a Dice coefficient similar to the intra-observer variability (0.91). The point-to-point error and point-to-mesh error were lower than the ones observed during the intra-observer reproducibility test, but that can be due to the fact that the \ac{RV} of the test/retest subject is smaller, and he is a pure control with better acoustic window than the American football players. Percent reproducibility of the region volumes are also similar to the ones observed in the intra-observer: we saw no suggestion that our method is more sensitive to inter-loop variability, but more samples are needed to validate. Unfortunately, at the moment of the study, we had no additional test/retest data available }.

\textcolor{black}{
We compared our method to registration. Both methods had similar reproducibility in the inlet and apical regions, but registration performed better in the \ac{RVOT} region. A known short-coming of registration is the depedence of the population used to build the atlas: we used remeshing to quantify the robustness of the methods to changes in the 3D-models processing and found that registration was more unstable to remeshing. This suggests that registration is less generalisable to be applied to meshes obtained from different segmentation models. In our synthetic dataset, our method was better for identifying circumferential apical and inlet growth, but worse in the \ac{RVOT} and at longitudinal elongations. However, elongation of the apex and \ac{RVOT} are not common in clinical cases, due to the constraints imposed by the pulmonary artery and pericardium.
}


\section{Conclusion}
\textcolor{black}{We proposed a geometry processing method for analysing the regional morphology of the \ac{RV} without depending on  point-to-point correspondence of image-based segmented meshes. The method parcellates the \ac{RV} in 3 regions: inlet, outlet and apical.} This parcellation also allowed to assess function via regional \ac{EF}. We analysed the reproducibility of the regional measurements, and found it below $8\%$ in both apex and  inlet segments for the intraobserver, but above $12\%$ in the interobserver case and outflow. Given that most of the instability comes from segmentation errors in the outflow portion, the addition of extra landmarks (like for example in the middle of the septum) would allow to improve reproducibility. 
We also proposed and used a novel method to generate localised remodelling patterns. We used it to generate synthetic remodelling surfaces to validate our parcellation method and showed that it captures global scaling of the ventricles as well as localised remodelling in the circumferential directions, but has difficulties in local longitudinal elongations.

\textcolor{black}{In future work, the use of this method to characterise different clinical populations involving dilatation of the \ac{RV} (such as tricuspid  regurgitation, pulmonary artery hypertension and arrhythmogenic right ventricular outflow tract) might allow to better characterise and understand their pathological remodelling.}

\section*{Acknowledgements}
This study was partially supported by the Spanish Ministry of Economy and Competitiveness (Maria de Maeztu Units of Excellence Programme - MDM-2015-0502), the European Union under the Horizon 2020 Programme for Research, Innovation (grant agreement No. 642676 CardioFunXion). \textcolor{black}{We thank doctors Duchateau and Nu{\~n}ez-Garc\'ia for fruitful discussions.}
